\documentstyle[preprint,aps]{revtex}
\begin{document}
\draft
\title{The Effective Theory of Hole Doped Spin-1 Chain
}
\author{Xi Dai, Bin Chen and Zhao-Bin Su}
\address{
Institute of Theoretical Physics, Academia Sinica,
P.O. Box 2735, Beijing 100080, China
}
\maketitle

\begin{abstract}
An effective theory for the hole doped spin-1 antiferromagnetic chain
is proposed in this paper. The two branches of low energy 
quasipaticle excitation
is obtained by the diagrammic technique. In the large t limit(in which
t is the hole hopping term), the lower band is essentially 
the bound state of one hole
and one magnon and the other band is the sigle hole state. 
We find a critical value of t, $t_c=0.21\Delta_H$(in which $\Delta_H$
is the Haldane gap).For $t>t_c$,
with the decrement
of t, the mixing of these two bands become stronger and stronger, and at
the same time the effective band mass becomes larger and larger. When
$t<t_c$ the minimum of the lower band moves away from zero the another
point between zero and $\pi/2$. The spin strcture factor is also calculated
in this paper, and we find that for large t limit the main contribution is
from the inter-band transition which induce a resonant peak in the 
Haldane gap. While for small t limit the main contribution is from the 
intra-band transition which only cause a diffusion like broad bump
in the Haldan gap.
\end{abstract}

\vspace*{.8cm}

{\bf Contact Author}: Xi Dai

{\bf Mailing Address}: Institute of Theoretical Physics,
 P.O. Box 2735, Beijing 100080,\\
\hspace*{4cm}  P. R. of China

{\bf E-mail}: daix@sun.itp.ac.cn

\newpage

\centerline{\bf I. INTRODUCTION}
\vspace{.5cm}

Since Haldane's seminal paper on the quantum Heisenberg spin chains\cite{Hal}, 
it has  been investigated by many theoretical and 
experimental studies. For
spin-1 antiferromagnetic chain, it is now well established both 
experimentally\cite{exp} and theoretically\cite{th1,th2}
 that there is an energy gap between 
the triplet excitation and the singlet ground state. The gap size
is recently determined to be $0.41J$ by using density matrix
renormalization group(DMRG)\cite{DMRG} and 
exact diagonalization\cite{ext}. It is known that the 
low energy behavior for the quantum
antiferromagnitc spin chains can be described by the nonlinear sigma
$(NL\sigma)$models\cite{Hal,NLS} and additional toplogical terms are further
 needed for half integer spin chains.
Particular for the spin-1 chain,
 a phenomenological model named as free boson
model has been proposed by Ian Affleck {\it et al}\cite{FB1,FB2,FB3}
 based on the large N
expansion for the $NL\sigma$ model. It has been shown that the free 
boson model captures the basic physics of the spin-1 chain in 
low temperature and can be used to calculate many physical 
properties correctly.  

There are mainly two families of 
compounds which exhibit the essential physics of
 the antiferromagnetic quantum spin-1 chain.
One is the early discovered compound $CsNiCl_3$\cite{exp}
 and the other
is $Y_2BaNiO_5$ . Replacing $Y$ 
by $Ca$ for $Y_2BaNiO_5$\cite{exp2,exp3}
 one can dope the spin-1 chains
by holes. The electronic
transport properties, polarized x-ray absorption and neutron 
scattering of $Y_{2-x}Ca_xBaNiO_5$ have been 
measured by J. F. DiTusa {\it et al}\cite{dop}.
 The result of electronic resistivity shows that the 
hole doping greatly reduces the resistivity which implies
a considerably large mobility for the holes. The neutron 
scattering experiment confirms the existence of new states in
the Haldane gap\cite{dop}. 
Theoretically the effect
of doping in Haldane gap system is a very interesting problem
and have been studied by several groups using both 
numerical\cite{st1} and analytical methods\cite{st2,st3}.

For case of static hole doping, by using the DMRG method\cite{st1} 
Sorensen and Ian Affleck studied the 
in-gap state caused by two kinds of doping. One is bond doping
and the other is  site doping. In bond doping case, the effect
of doping is to perturbe an antiferromagnetic bond with $J'\neq J$ .
It gives rise 
a localized state centered at the perturbed bond with
 a descrete energy level appeared inside the Haldane gap for sufficiently
strong or weak $J'$. In site doping case,  a static hole is 
considered to be located on the O ion between two Ni ions. Therefore
the super exchange process between these two spins is destroyed 
or partly destroyed by the hole.
The bound states in the Haldane gap are found only when the 
coupling between the hole and the nearest spin is weaker than
 a critical value. Similar 
results were  obtained by M.Kaburagi and {\it et al} using variational
calculation\cite{st3}.

Also one can consider a mobile hole doped in spin-1 chain.
S.C. Zhang and D. P. Arovas\cite{zhang} considered a spin-0 hole hopping in
a background of spin-1 chain by using an effective model which is quite
similar to the t-J model and found some exact single and multi-
hole spin singlet wave functions. The problem of spin 1/2 hole 
moving in spin-1 chain has been considered
 by K. Penc and H. Shiba\cite{penc} 
in the limit of small hopping amplitude. In their approach,the holes
are hopping between the O sites and destroy the corresponding 
super exchange 
processes completely when the O sites are occupied by the holes.
They found
one spin-3/2 band and two spin 1/2 bands either in the
VBS model or in the Heisenberg model due to the interaction 
between hole and its nearest neighboring spins. 
But for the realistic 
Heisenberg model, their variational calculation
can only treat the finite lattice up to 15 sites. Recently E. Dagotto  
proposed a t-J like model to study
the hole doped spin-1 chain. Using numerical techniques\cite{Dag}, the 
dispersion of the effective hole bands and the spin structure
factor were obtained for the finite lattice up to 12 sites.

In the present paper, we propose an effective continuum theory
to study the hole doping in spin-1 chain. The spin-1 chain is
modeled by the free Boson model proposed by Ian Affleck et 
al\cite{FB1,FB2,FB3}
which is essentially derived from the large N expansion of the $NL\sigma$ model. The
effective interaction between the holes and the magnons are 
derived, based upon the following considerations.
Recall the static hole limit, the effective interaction only
contains the scattering process and the holes act as the
scatters of the magnons. Then the effective Hamiltonian for
the static hole doped in spin-1 chain can be even easily intepreted
in the first quantizaction picture. The result shows that there
are one or two (depends on the inetraction strength)
impurity states with total spin equals $1/2$ or $3/2$ 
respectively in the Haldane gap which are actually the bound
states of a hole and a spin-1 magnon. This result is in good
agreement with the numerical results.We may then draw an intuition
to the moving holes. When the hole 
moves,the effective interaction should contains two main terms, one
describes the scattering process of the hole and the magnon which
has the same origin with that of the static hole case,
the other describes the emission and absorption process
of magnons.  

We show in this paper that the first term
results in a bound state of one hole and one magnon with the 
bound energy strongly depending on the total spin of the 
hole and magnon. Therefore, we will find that if we only consider
the scattering process, the result is very similar to the static
hole case. The only difference is that each impurity state in the static
hole case will extend to a corresponding energy band if the hole can hop. 
The effect of the emission and asorption
term will result a hybridization of  single hole state 
and the bound state of one
hole and one magnon with total spin 1/2. 
This term  plays a crucial roll for small t, 
because the energy level of the bound state and that of the 
single hole state is very close in small t limit and is quite
large in large t limit. Then we obtain  three branches of 
excitation, the bound state of one hole and one magnon with
total spin 3/2, the ``one particle like'' state with total
spin 1/2 and the ``two particle'' like state with total spin
1/2. The ``one particle like'' state approaches to the single
hole state in large t limit and the ``two particle like''
state approaches to the bound state of one hole and one
magnon with total spin 1/2.The dispersion of these three branches
of excitations are obtained in the whole range of the 
hopping amplitude t.  The band minmum of the ``two particle like''
state is found to be located 
at $\pi$ for $t>t_c$
and will be move toward $\pi/2$ when $t<t_c$. The value of $t_c$
is found to be near $0.21\Delta_H$ in our calculation.
In fact this can be understood as a result of the above
mentioned hybridization effect in the small t case.
 These results
are consistent with the results for VBS model obtained by
 K. Penc and H. Shiba. 

The spin 
structure factor is also obtained and the result is quite
different for large t case and small t case. For large t case 
the hole contribution to the  spin structure factor is mainly contributed
from the interband transition(from ``two particle like'' state
to ``one particle like'' states) and the spectral weight whithin 
the Haldane gap is centered at the minimum energy cost between the two
bands, which is consistent with the calculation of Dagotta {\it et al}.
 But for small t case the contribution is mainly from
intra band transition(from ``two particle like'' state to
``two particle like'' state within the same branch of dispersion),
 so there exists a diffusion like broad bump in 
the Haldane gap. The difference of the spin structure factor
in the above mentioned two limit is quite interesting and has never
been mentioned in the previous works.

The rest of the paper is organized as follows: 
The effective continuum Hamiltonian is derived in Sec. II, 
whereas dispersion of the bound states  is claculated in Sec. III,
and the spin structure factor is calculated in Sec. IV.
Finally, we make a few concluding remarks in Sec. VI.

\bigskip
\centerline{\bf II. The Derivation of The Effective Hamiltonain}
\vspace{.5cm} 

In this paper, we assume that the holes can hop whithin the oxygen
sublattice, which is shown in Fig.1. 

Then we can  begin with the following total Hamiltonian
$$
H_{total}=H_{ch}+H_{h}+H_{J'}+H_{J_1}+H_{J_2}
$$
$H_{ch}$ is the rotationally invariant spin Hamiltonian for
Antiferromagnetic spin-1 chain.
\begin{mathletters}
\begin{equation}
H_{ch}=J\sum_i {\vec S_i} \cdot {\vec S_{i+1}}
\end{equation}
and $H_h$ is the hopping term of holes
\begin{equation}
H_h=-t\sum_n f_{n+1/2\sigma}^+f_{n+3/2,\sigma}+H.C.
\end{equation}

$H_{J'}$ and $H_{J_1}$ represent the destroying of 
super exchange process and a Kondo like interaction of spins
and holes respectively,
\begin{equation}
H_{J'}=-J'\sum_{\sigma,n} {\vec S}_n\cdot {\vec S}_{n+1} 
f_{n+1/2,\sigma}^+f_{n+1/2,\sigma}
\end{equation}
\begin{equation}
H_{J_1}=J_1\sum_{n,\alpha,\beta}{\vec \sigma}_{n,\alpha,\beta}\cdot
(\vec S_n+\vec S_{n+1})f_{\alpha,n+1/2}^+f_{\beta,n+1/2}
\end{equation}
And $H_{J_2}$ describes the hole hop to another
site with its spin flipped by interacting with the spins of the chain.
\begin{equation}
H_{J_2}=J_2\sum_{n,\alpha,\beta}{\vec \sigma}_{n,\alpha,\beta}\cdot
\vec S_n f_{\alpha,n-1/2}^+f_{\beta,n+1/2}+H.C.
\end{equation}
\end{mathletters}

It is known that $H_{ch}$ can be mapped to the non-linear $\sigma$
model by using the path intergral in the 
spin coherent state representation. The antiferromagnetic order parameter is
represented by a three dimentional vector field $\vec \phi$\cite{NLS},
obeying a constraint as $ |\vec \phi|^2=1 $. The uniform 
magnetisation is represented by $\vec l$,

\begin{mathletters}
\begin{equation}
{\vec l}=({1 \over vg}){\vec \phi}\times{\partial {\vec \phi}
\over {\partial t}}
\end{equation} 
And the spin operator at site i can be written as,
\begin{equation}
{\vec S_i}\approx s(-1)^{i}{\vec \phi}+{\vec l},
\end{equation}
Then the spin-1 Heisenberg Halitonian can be mapped into an effective
continuum field theory with the Hamiltonian  density as:

\begin{equation}
{\cal H}_{ch}={{v}\over 2} {\vec l}^2+{v\over {2}}\left ({\partial{\vec \phi}}
\over {\partial x} \right )^2
\end{equation}
\end{mathletters}
with the constraint $ |\vec \phi|^2=1 $.

By taking the large N(the number of the components for ${\vec\phi}$
field) limit, and further
introduce a mass term as the Lagrange multiplier
to relax the constraint on the field $\vec \phi$ into an averaged one,  
we follow 
the free Boson model or Ginsburg-Landau model proposed
by I. Affleck {\it et al}\cite{FB1,FB2,FB3},in which an
additional ${\vec \phi}^4$ term is added for keeping the stability.
We then have
\begin{equation}
{\cal H}_{ch}= {{v}\over 2} {\vec \Pi}^2+{v\over {2}}\left ({\partial{\vec \phi}}
\over {\partial x} \right )^2+{1\over{2v}}\Delta_H^2\phi^2+
\lambda |{\vec \phi}|^4
\end{equation}
where ${\vec \Pi}(x)$ is the canonic momentum conjugated to the field
${\vec \phi}(x)$ satisfying $[\phi(x)_{\alpha},\Pi(x')_{\beta}]
=\delta_{\alpha,\beta}\delta(x-x')$.
The three parameters $v$, $\Delta_H$ and $\lambda$ are chosen 
phenomenologically to fit the experiment or the numerical result.
In our present study we omit the ${\vec \phi}^4$ term as for a 
preliminary discussion. We can then diagonalize the ${\cal H}_{ch}$
 by Fourier
transformation, and obtain three branches of free magnons, which 
describe the triplet excitations upon the singlet ground state.
$$
H_{ch}={{v}\over 2}\sum_k{\vec \Pi}(-k)\cdot{\vec \Pi}(k)+
[{{vk^2}\over 2}+{{\Delta_H}^2\over {2v}}]{\vec \phi}(-k)\cdot{\vec \phi}(k)
$$
\begin{equation}
=\sum_{k,r} E_k (a_{k,r}^+a_{k,r}+{1\over 2})
\end{equation}
In equation(4) $a_{k,r}$ $a_{k,r}^+$ are the 
annihilation and creation
operators of the magnons  satisfying
$$
[a_{k,r},a_{k',r'}^+]=\delta_{rr'}\delta_{kk'}
$$
with
$r=x,y,z$, $E(k)=\sqrt {v^2k^2+\Delta_H^2}$ and:
\begin{mathletters}
\begin{equation}
{\vec \phi}(k)=\sqrt{v\over 2E_k}\left (\vec a_k+\vec a_{-k}^+ \right )
\end{equation}
\begin{equation}
{\vec \Pi}(k)=i\sqrt{E_k\over 2v}\left (\vec a_k-\vec a_{-k}^+ \right )
\end{equation}
\end{mathletters}

Now we have expressed the Hamiltonian for the Heisenberg chain in terms of
the three branches of gapful magnons, we have
In the continuum limit,
\begin{equation}
({\vec S}_i+{\vec S}_{i+1})\approx 2{\vec l}(x_i)+
{{\partial {\vec \phi}}\over {\partial x}}|_{x=x_i}
\end{equation}
For the perfect
spin-1 chain, the field ${\vec l}(x)$ is always one order
smaller than the field $\vec\phi(x)$ and is of the same order as that of
${\partial{\vec\phi(x)}}\over{\partial x}$ which reflects the short
range anti-ferromgnetic correlation in low temperature. But if doped with
an spin-1/2 holes, as we will show later, the hole will induce
a localized mode of magnon which has an extension only of
several lattice near 
the hole. 
So in the doped case near the site of the hole, the field
${\vec l}(x)$ could be of the same order with field 
$\vec\phi(x)$, and could be  one order larger than
${\partial{\vec\phi(x)}}\over{\partial x}$.
Therefore, we can omit ${\partial{\vec\phi(x)}}\over{\partial x}$
in (6) and only keep the first term.
Then the interaction term $H_{J'}$, $H_{J_1}$ and $H_{J_2}$ can be written
as,
\begin{mathletters}
\begin{equation}
H_{J'}=-{J'\over 2}\sum_i\left [(\vec{S_i}+\vec{S_{i+1}})^2-4\right ]
f^+_{i+1/2,\sigma}f_{i+1/2,\sigma}
\approx-2J'\sum_i\left [ \vec{l}(x_i)^2-4\right ]
f^+_{i+1/2,\sigma}f_{i+1/2,\sigma}
\end{equation}

\begin{equation}
H_{J_1}=2J_1\sum_i \left [\vec{l}(x_i)+
{\partial{\vec \phi} \over \partial x} \right ] 
\cdot {\vec \sigma}_{\alpha,\beta} 
f^+_{i+1/2,\alpha}f_{i+1/2,\beta}
\approx 2J_1\sum_i \vec{l}(x_i)\cdot {\vec\sigma}_{\alpha,\beta} 
f^+_{i+1/2,\alpha}f_{i+1/2,\beta}
\end{equation}

\begin{equation}
H_{J_2}=J_2\sum_i \left [\vec{l}(x_i)+{\vec \phi}(x_i)\right ]
\cdot {\vec \sigma}_{\alpha,\beta} 
f^+_{i+1/2,\alpha}f_{i-1/2,\beta}
\end{equation}
\end{mathletters}
We can now express the field $\vec{l}(x_i)$,  $\vec{l}(x_i)^2$
and ${\vec \phi}(x_i)$ in terms of 
 the magnon creation and annihilation
operators. We leave the detail derivation
 in appendix A. After discarding
the multi-magnon processes, the Hamiltonian reads,

\begin{equation}
H_{ch}=\sum_{k,\mu} E_k (a_{k,\mu}^+a_{k,\mu}+{1\over 2})
\end{equation}
in which $\mu=-1,0,1$ and
\begin{mathletters}
\begin{equation}
a^+_{\pm}(k)=\mp{1 \over \sqrt{2}}\left [a_x(k)^+\pm ia_y^+(k)\right ]
\end{equation}
\begin{equation}
a_0^+(k)=a_z^+(k)
\end{equation}
\end{mathletters}
\begin{equation}
H_{J'}=-4\sum_{i,q,k,k',\mu,\sigma}\gamma'(k',q)J'a^+_{\mu,k'+q}a_{\mu,k'}
f^+_{\sigma,k-q}f_{\sigma,k}
\end{equation}
\begin{equation}
H_{J_1}=\sum_{\mu\nu,k,k',q,\alpha,\beta}
2\gamma_1(k',q) J_1
{\vec S}_{\mu\nu}\cdot {\vec \sigma}_{\alpha,\beta}
a^+_{\mu,k'+q}a_{\nu,k'}f^+_{\alpha,k-q}f_{\beta,k}
\end{equation}
as well as
$$
H_{J_2}=\sum_{\mu\nu,k,k',q,\alpha,\beta}2\gamma_1(k',q)
J_2cos(k){\vec S}_{\mu\nu}\cdot {\vec \sigma}_{\alpha,\beta}
a^+_{\mu,k'+q}a_{\nu,k'}f^+_{\alpha,k-q}f_{\beta,k}
$$
\begin{equation}
+\sum_{\mu\nu,k,q,\alpha,\beta}
2J_2\sin(k)\sqrt{v\over\Delta_H}{\sqrt{3}\over 2}
D_{\mu\alpha,{1 \over 2}\beta}
a_{\mu}^+(-q)f^+_{\alpha}(k+q-\pi)f_{\beta}(k)+H.C.
\end{equation}
in which $D_{\mu\alpha,{1 \over 2}\beta}=<\mu\alpha|{1\over 2}\beta>$
is the transformation matrix between the $(J^2,J_z)$ representation and 
that of $(S_z,\sigma_z)$.

In this paper, we choose the amplitude of the Haldane gap $\Delta_H$ as the
unit of energy, so we have $\Delta_H=1$. We choose the value of $v$ 
by fitting the correlation length at zero temperature, which is 
believed to be closed to 7\cite{penc}. So we have
$\xi={v \over \Delta_H}=7$, then we have $v=7$.
The two vertex functions $\gamma_1(k,q)$ and $\gamma'(k,q)$
are derived in appendix A. It is quite difficult for us to treat this
k,q dependent vertex function analytically. So as a low energy effective
theory, we simply replace the two k,q dependent vertex function by
two k,q independent phenomenological parameters $\lambda_1$ and
$\lambda'$ respectively, whose value can be fixed by fitting
our approach to the 
existing numerical results for the static hole problem.

In the static limit $t=J_2=0$, the effective Hamiltonian is
quite simple especially in the continuum limit,
$$
H=\int dx \sum_{\mu}a_{\mu}^+(x)(1-{1\over {m_b}}{\partial^2\over
{\partial x}^2})a_{\mu}(x)
$$
\begin{equation}
-4\lambda'J'\sum_{\gamma}
a_{\gamma}^+(0)a_{\gamma}(0)+\sum_{\gamma\mu}
2\lambda_1J_1{\vec S}_{\gamma\mu}
\cdot {\vec s}_{imp} a_{\gamma}^+(0)a_{\mu}(0)
\end{equation}
in which
\begin{equation}
a_{\mu}(x)={1 \over \sqrt{N}}\int_{-\infty}
^{\infty}dk a_{\mu}(k)e^{-ikx}
\end{equation} 
To study the low energy excitation inside the Haldane gap, we need
 only
consider the one magnon state. And this will lead to the following
schrodinger equation in first quantized picture.
\begin{equation}
\left [-{1\over{2m_b}}{\partial^2\over{\partial x^2}}+
2\lambda_1J_1{\vec s}_{imp}
\cdot {\vec S}\delta(x)-4\lambda_2J'\delta(x)+\Delta_H \right ]\psi(x)
=E\psi(x)
\end{equation}
The magnon wave function $ \psi(x)$ has six components corresponding
to six eigen states of $s^z_{imp}$ and ${S^z}$. We can easily solve
the above  schrodinger equation by first diagnolizing Hamiltonian in
spin space. We have
\begin{equation}
\left [-{1\over{2m_b}}{\partial^2\over{\partial x^2}}+
2J_1\lambda_1 e_j\delta(x)-4\lambda_2J'\delta(x)+\Delta_H \right ]\psi_j(x)
=E\psi_j(x)
\end{equation} 
in which j is the total spin of the impurity and the magnon, and
$e_j=j(j+1)/2-(S^2+s_{imp}^2)/2$, so
\begin{equation}
e_{1/2}=-\Delta_H~~~~~~~~~~~~~~~~~~~e_{3/2}=1/2\Delta_H
\end{equation}
The Eq.(14) describes a particle with mass $m_b$ bounded by a $\delta$
potential which can be solved easily. The corresponding eigenvalues
and eigen functions are:
\begin{mathletters}
\begin{equation}
E_{1/2}=\Delta_H-2m_b(\lambda_1J_1+2\lambda_2J')^2~~~~~~
\psi_{1/2}(x)={1\over\sqrt{L}}
\exp^{-|x|/L}~~~~~~with~~~L={1\over{2m_b(\lambda_1J_1+2\lambda_2J')}}
\end{equation}
\begin{equation}
E_{3/2}=\Delta-2m_b(-0.5\lambda_1J_1+2\lambda_2J')^2~~~~~~
\psi_{3/2}(x)={1\over\sqrt{L}}
\exp^{-|x|/L}~~~~~~with~~~L={1\over{m_b(-\lambda_1J_1+4\lambda_2J')}}
\end{equation}
\end{mathletters}
Eq.(18.b) is only meaningful for case of $\lambda_1J_1<4\lambda_2J'$
If $\lambda_1J_1>4\lambda_2J'$ 
the effective potential for
the state with total spin 3/2 is repulsive which can be easily 
verified from eq.(16). Therefore there is 
no bound state with total spin 3/2 in this case. 
When $\lambda_1J_1<4\lambda_2J'$
there always exist two bound state where 
one has two-fold degeneracy
and the other has four-fold degeneracy. We can compare our results
with the corresponding 
numerical results. Firstly we choose $J'=J=2.5\Delta_H$ 
which corresponds to the case that
the super exchange interaction got destroyed entirely. 
If we have further $J_1=0$, the problem becomes precisely
equivelent to an open chain. The numerical studies show that the 
ground state of the open spin-1 chain is four-fold\cite{open1,open2}. 
In the simple approach engaged
in this paper,
it gives a nice description of these four-fold states. If we simply
choose $\lambda_2=1.01$ the energy
cost to create one localized magnon near the chain edge
 is 0 determined by (18.b), then
the degenerated ground state will be zero magnon state as well as three-
fold states with one localized magnon near the edge, which is altogether
four-fold.
The edge states
decribed by (18.b) extend to 5 lattice which is also 
consistent with the numerical result. For finite $J_1$, equation 
(18.a) leads
to $E_{1/2}\approx -0.4\lambda J_1,
E_{3/2}\approx 0.2\lambda J_1$ in small $J_1$ 
limit while the numerical
study predicts that $E_{1/2}\approx -J_1,E_{3/2}\approx 0.5J_1$.
We then choose $\lambda_1=2.5$ to fit the numerical results.   
Fig.2 shows the $-E_{1/2}$ as the function of $J_1$ in a 
full range of $J_1$. Compared with the results obtained by Ian Affleck
{\it et al}\cite{st1}, we find that our 
simple treatment fits quite well with the numerical
results espcially in weak coupling regime.

\bigskip
\newpage
\centerline{\bf III. The Dispersion of the Bound State of Magnon and Hole}
\vspace{.5cm} 
In this section we consider the bound state constituted by one
magnon and one hole. In the present paper, we assume that the local
super exchange is destroyed entirely, so we choose $J'=J$, 
$\lambda_1=2.5$, $\lambda'=1.01$,$J_1=J_2=0.2\Delta_H$,
 and $m_b=1/49$. We will consider
this problem in two limiting cases, one is the large t limit and the other
is small t limit. We can reorganize the total Hamiltonian as
$H=H_h+H_{ch}+H_1+H_2$, in which $H_h$ and $H_{ch}$ are the free
Hamiltonian for holes and spins respectively as shown in (1b) and (4)
and $H_1$ and $H_2$ are two kinds of interaction representing
the scattering process and the magnon emission and absorption process
respectively.
\begin{equation}
H_1=\sum_{\mu\nu,k,k',q,\alpha,\beta}
\left [ 2\lambda_1 J_1+2\lambda_1J_2cos(k)\right ]
{\vec S}_{\mu\nu}\cdot {\vec \sigma}_{\alpha,\beta}
a^+_{\mu,k'+q}a_{\nu,k'}f^+_{\alpha,k-q}f_{\beta,k}
\end{equation}

\begin{equation}
H_2=\sum_{\mu\nu,k,q,\alpha,\beta}
g(k)D_{\mu\alpha,{1 \over 2}\beta}
a_{\mu}^+(-q)f^+_{\alpha}(k+q-\pi)f_{\beta}(k)+H.C.
\end{equation}
in which $g(k)=2J_2\sin(k)\sqrt{v\over\Delta_H}{\sqrt{3}\over 2}$. 

\bigskip
\noindent{\bf a. The large t limit}

In latge t limit, the holes are mainly distributed near $P=0$,
so we can treat the bare hole dispersion in continuum approximation.
We have $E_h(P)=-2t(cosP-1)\approx {p^2\over {2m_f}}$, in which
$m_f={1\over 2t}$.

In order to study the dispersion relation
of the bound state, we should consider the two-particle Green's
function with total spin j(j=1/2 or 3/2) and its z-component m
which is defined as the following.
\begin{equation}
\Gamma_{jm}(x,t)=-i\sum_{\alpha'\mu'\alpha\mu}<jm|\alpha'\mu'>
<{\cal T}a_{\mu'}(x,t)f_{\alpha'}(x,t)|a_{\mu}^+(0,0)f_{\alpha}^+(0,0)>
<\alpha\mu|jm>
\end{equation}
The above two-particle Green's function can be obtained by using
Fenyman diagram expansion.
One may easily verify that the magnon emission and absorption
 term only act on the total spin-1/2
state and can't affect the state with total spin 3/2. So the spin-3/2
state only feels the scattering term as shown in Fig.3, which 
leads to,
\begin{equation}
\Gamma_{3\over 2}(P,i\nu)=\pi_0(P,i\nu)+
{\tilde\pi}_0(P,i\nu)
\Gamma_{3\over 2}(P,i\nu)
\end{equation}
So we have
$$
\Gamma_{3\over 2}(P,i\nu)=
{{\pi_0(P,i\nu)} \over {1-{\tilde\pi}_0(P,i\nu)}}
$$

\begin{equation}
{\tilde\pi}_0(P,i\nu)
=\int_{-\pi}^{\pi}dq V_{3\over 2}(P-q)
 {{1+n_B({q^2\over{2m_b}}+1)-n_F({(P-q)^2\over{2m_f}})}
\over {i\nu-{(P-q)^2\over{2m_f}}-{q^2\over{2m_b}}-\Delta_H}}
\end{equation}
and
\begin{equation}
\pi_0(P,i\nu)
=\int dq{{1+n_B({q^2\over{2m_b}}+1)-n_F({(P-q)^2\over{2m_f}})}
\over {i\nu-{(P-q)^2\over{2m_f}}-{q^2\over{2m_b}}-\Delta_H}}
\end{equation}
in which $V_{3\over 2}(k)=\lambda_1\left [J_1+J_2cos(k)\right ]
-4\lambda_2J'$ and P is the totale momentum for
 the hole-magnon system. 

In above equations $n_B$ and $n_F$ are the Boson and Fermi 
distribution function respectively. But in our present 
study if we consider only the dilute limit of the hole
doping, so we can assume that the temperature satisfy 
$T_F\ll T \ll \Delta_H$, in which $T_F$ is the Fermi temperature of the
holes. Then we can replace the Fermi-Dirac distribution function
$n_F(E)$ by its classic limit which is the Boltzman distribution
function. In equation(26), we can further ignore the $n_B$ and
$n_F$ in dilute doping and low temperature case.Then we have
\begin{equation}
{\tilde\pi}_0(P,\omega+i0^+)\approx   
\int dq {{V_{3\over 2}(P-q)}
\over {\omega+i0^+-{(P-q)^2\over{2m_f}}-{q^2\over{2m_b}}-Delta_H}}
=\int_{-\pi}^{\pi}dq'{ {V_{3\over 2}[(1-{m_b\over M})P-q']}
\over {\omega+i0^+-{P^2\over{2M}}-{q'^2\over{2\mu}}-\Delta_H}}
\end{equation}
in which $q'=q-{m_b\over M}P$ and ${1\over \mu}={1\over m_f}+{1\over m_b}$.
 We may  expand the functiom
$V_{3\over 2}$ with respect to $q'$ and only keep the leading order, because
the main contribution to the integration comes from
$q'<\sqrt{2m_b(\Delta_H+{P^2\over {2M}}-\omega)}$, which is quite 
small in the $\omega$ regime  in which we are interested. 
So we have
\begin{equation}
Re{\tilde\pi}_0(P,\omega+i0^+)\approx
-{1\over 2} {V_{3\over 2}[(1-{m_b\over M})P]}\sqrt{{2\mu}\over
{Delta_H-\omega+{P^2\over{2M}}}}
\end{equation}
and
\begin{equation}
Re\pi_0(P,\omega)=-{1\over 2}\sqrt{{2\mu}\over
{\Delta_H-\omega+{P^2\over{2M}}}}
\end{equation}
for $\omega<\Delta_H$.

The bound energy for the bound state of one hole and one magnon
with total spin-3/2 is determined by the pole of 
$\Gamma_{3\over 2}(P,\omega)$, which is 
$1-Re{\tilde\pi}_0(P,\omega)=0$. We then have 
\begin{equation}
E_{3/2}(P)
=1-{1\over 2}\mu V_{3/2}\left [(1-{m_b\over M})P\right ]^2
+{P^2\over {2M}}
\end{equation}

The situation of the bound state with total spin-1/2 is
not as simple as the case of total spin-3/2, because the 
interaction term $H_{J_2}$ will mix the two particle state
and the one particle state, which is illustrated in Fig.3.
Therefore we should take into consideration the contribution
from both the two terms. Following the diagrammatic rule 
shown in Fig.(4b), we then have

$$
\Gamma(P,i\nu)=\Gamma_0(P,i\nu)+
\Gamma_1(P,i\nu)G(P+\pi,i\nu)\Gamma_1(P,i\nu)+
$$
\begin{equation}
\Gamma_1(P,i\nu)G(P+\pi,i\nu)\Gamma_2(P,i\nu)G(P+\pi,i\nu)
\Gamma_1(P,i\nu)+...
\end{equation}
where
\begin{equation}
\Gamma_0(P,i\nu)
={\pi_0(P,i\nu) \over {1-V_{1\over 2}\left [(1-{m_b\over M})P\right ]
\pi_0(P,i\nu)}}
\end{equation}
\begin{equation}
\Gamma_1(P,i\nu)
={ g(P,{m_b\over M}P)\pi_0(P,i\nu) 
\over{1-V_{1\over 2}\left [(1-{m_b\over M})P\right ]
\pi_0(P,i\nu)} }
\end{equation}
and
\begin{equation}
\Gamma_2(P,i\nu)
= g(P,{m_b\over M}P)^2\pi_0(P,i\nu)+
{{ g(P,{m_b\over M}P)^2\pi_0^2(P,i\nu)
V_{1\over 2}\left [(1-{m_b\over M})P\right ]} 
\over{1-V_{1\over 2}\left [(1-{m_b\over M})P\right ]
\pi_0(P,i\nu)} }
\end{equation}
in which $V_{1\over 2}(k)=2\lambda_1\left [J_1+J_2cos(k)\right ]
-4\lambda'J'$.
To derive the above equation, the same approximation is done
as (26).Then we have
\begin{equation}
\Gamma(P,\omega)={\Gamma_0(P,\omega)\over{1-g^2(P,{m_b\over M}P)
G(P+\pi,\omega)\Gamma_0(P,\omega)}}
\end{equation}
in which 
\begin{equation}
G(P+\pi,\omega)={1\over {\omega-4t+{P^2/{2m_f}}}}
\end{equation}
is the Green's function near $\pi$. Then we can obtain the energy
of the bound state with total spin-1/2 by solving the eqaution:
\begin{equation}
{1-g^2(P,{m_b\over M}P)
G(P+\pi,\omega)\Gamma_0(P,\omega)}=0
\end{equation}
The result is 
\begin{equation}
E_{1\over 2}(P)=\Delta_H-{1\over 2}\mu V_{1\over 2}^2+{P^2\over{2M^*}}
\end{equation}
with the renormalized effective mass
\begin{equation}
{1\over M^*}={1\over M}-{{3J_2^2(1-{m_b\over{2M}})^2\xi\mu V_{1\over 2}(0)}
\over {{1\over 2}(0)\mu V_{1\over 2}^2-\Delta+{2\over m_f}}}
+4*\mu\lambda_1J_2\left [\lambda_1(J_1+J_2)+2J'\lambda' \right ]
\end{equation}
in which $m_f={1\over {2t}}$.
We find that if t is smaller than a critical value $t_c$, the
effective mass will change sign which means the band minimum may
move away from the $P=0$ state
and will be located at a point between 0 and $\pi$. From the above
expression, the $t_c$ is determined by the condition ${1\over M_*}=0$.
And the result is $t_c=0.21\Delta_H$.

For the case of $t>>t_c$, the magnon emission and absorption
processes are no longer important and can be ignored, so the
number of the magnons is conserved. Then the essential physics
 can be viewed more clearly
as a two-body problem in the first quantization picture.
We can write again a Schrodinger equation for
the hole-magnon system.

\begin{equation}
\left [-{1\over {2m_b}}{\partial^2\over{\partial x_b}^2}-
{1\over {2m_f}}{\partial^2\over{\partial x_f}^2}+
{\bar J}{\vec s_{imp}}\cdot {\vec S}\delta(x_b-x_f)-
4\lambda_2J'\delta(x_b-x_f)+1\right ]\psi(x_b,x_f)=E\psi(x_b,x_f)
\end{equation}
in which $m_f={1\over {2t}}$ and ${\bar J}=2\lambda_1(J_1+J_2)$.
The above Halmiltonian describes a two body problem with an 
attractive interaction between them.
We can divide the Hamiltonian into two parts, one describe the 
motion of center of mass and the other describe the relative
motion.
\begin{mathletters}
\begin{equation}
H_c=-{1\over {2M}}{\partial^2\over{\partial {\bar x}}^2}
\end{equation}
\begin{equation}
H_r=-{1\over{2\mu}}{\partial^2\over{\partial x^2}}+
{\bar J}{\vec s}_{imp}
\cdot {\vec S}\delta(x)-4\lambda_2J'\delta(x)+1 
\end{equation}
\end{mathletters}
with $\mu={m_fm_b\over{m_f+m_b}}$,$M=m_b+m_f$,$x=x_f-x_b$ and
$\bar x={{m_bx_b+m_fx_f}\over M}$. Equation (23a) describes a
free propogation of a composite particle constituted by a hole 
and a magnon  with total mass $M$, and equation (23b) describes
a particle in a attractive potential which is solved in (19a) and
(19b). So the situation in large t limit is quite clear. 
One magnon and one hole form a bound state with the bound energy 
$1-2\mu(-0.5\lambda_1J_1+2\lambda_2J')^2$
for $S_{total}=3/2$ state or a bound energy 
$1-2\mu(\lambda_1J_1+2\lambda_2J')^2$ for
$S_{total}=1/2$ state. These results are consistent with and further confirm
the results obtained by using Feynman diagram techeniques
in the case of $t>>t_c$.

 The center of mass moves like a free particle
with total mass $M$.
So in the case of large t limit, the bound state of magnon 
and hole with total spin 1/2 is energetically favorable
against the free hole state.Therefore, the ground state
of the system will be the bound pairs of the holes and magnons with
total spin 1/2, in another words, every hole will induce a magnon
binding with it.
Further, these bound pairs can be viewed as a composite Fermions
with spin 1/2 which can propogate freely in the spin-1 chain.

\bigskip
\noindent{\bf b. The Small t limit}
\bigskip

In the small t limit, we should calculate the same diagrams as in
large t limit.
Also we can calculate the dressed two particle Green's function
by considering the diagramm shown in Fig(4). 
 The difference between the two limiting cases is that we can't
use the long wave length expansion anymore for holes in the small
t case.   
The $\pi_0(P,\omega+10^+)$ in small t limit can be 
written as:
\begin{equation}
\pi_0(P,\omega+i0^+)=   
\int dq {1
\over {\omega+i0^++(2t\cos(P-q)-2t)-{q^2\over{2m_b}}-\Delta_H}}
\end{equation}
Since in small t limit we always have $1/m_b\gg t$, the main 
contribution for the integrand comes from a quite narrow
q regime near q equals to zero. So we can
 expand $\cos(P-q)$ in q near P, 
that is
\begin{equation}
\pi_0(P,\omega+i0^+)=   
\int dq {1
\over {\omega+i0^++2t\cos(P)+2t\sin(P)\cdot q-t\cos(P)q^2
-{q^2\over{2m_b}}-\Delta_H-2t}}
\end{equation}
then intergrate it using the method described before.
Also we can obtain the values of $\pi_1$ and $\pi_2$ using
the same method. The dispersion relation of the qusiparticle which
is the mixed state of the single hole state and the bound state
of one magnon and one hole is determined by searching for the low energy
poles of the Green's function similar to what we have done
 in the large t limit. 
Two
branches of excitation are obtained with energy $\omega_{\pm}(P)$
as shown in Fig.5 for various value of t. Near the poles, the two
particle propagator can be written as
\begin{equation}
\Gamma(P,\omega+i0^+)={z_i(P)\over {\omega-\omega_i+i0^+}}
\end{equation}
for $i=+,-$. The result of $z_i(P)$ is shown in Fig.6.
From Fig.6 we can find  that for P near zero 
the lower branch is almost two
particle like and the higher branch is one particle like,
but for P near $\pi$ the lower branch is one particle like and
the higher branch is two particle like.  
Between the above two limit, the one particle state and two
particle state are strongly hybridized. So near $0$ and $\pi$, 
the mixing of the one particle state and the two particle 
state is quite small and the effect of the magnon emission
and absorption term is only modifying the effective mass
of the bounded states. But for P near $\pi/2$, the one particle
state and the two particle state are strongly hybridized, which 
may split the two energy level remarkably. When t is large this
hybridization effect
only increases the effective mass. But if t is sufficiently small
the energy split caused by $J_2$ term which is of the order
$J_2$  could be much large than the kinetic energy of order t.
Then the band minimum of the lower excitation will no longer stay 
at $P=0$ but move towards
$\pi/2$. We can find from Fig.5 the critical value of t is 
$0.21\Delta_H$, which is consistent with the value that we obtained 
in the large t limit calculation. From equ.(12) we can see that the origin of the
magnon emission and absorption term is the coupling between the
mobile holes and the antiferromagnetic field $\vec \phi(x)$. Although
the long range antiferromagnetic order is not stable
in 1D spin chain, the short range antiferromagnetic fluctation
is still quite strong in low temperature. If the holes can feel
such antiferromagnetic fluctation, the band minimum will move 
to $\pi/2$. It is rather interesting to note that
this feature is quite similar to that of the single hole
moving in the 2D antiferromagnetic lattice. 

 Also, we can calculate the renormalized single particle
 Green's function considering the diagrams showing in Fig(4c). 
And the same quasiparticle poles are obtained.But we provide here
a transparent intuitive picture. 

Our results for small t limit are very similar with the results
obtained by Shiba el al\cite{penc}  for the VBS model, in which the
band minimum is moved towards $\pi/2$ when the hopping term
$t$ is smaller than a critical value.

\bigskip
\bigskip
\centerline{\bf IV. Calculation of the Spin Structure Factor}
\vspace{.5cm} 

The spin structure factor can also be calculated in our
approach. Since in the doped spin-1 chain both the holes and the
spins can contribute to the spin structure factor, so we have
\begin{equation}
{\cal S}(\pi,\omega)={1\over 2\pi}\int dt e^{i\omega t} 
<({\vec\phi}(0,t)+\sum_k{\vec \sigma}_{\alpha\beta}f^+_{\alpha,k+\pi}(t)
f_{\beta,k}(t))|
({\vec\phi}(0,0)+\sum_k{\vec \sigma}_{\alpha\beta}f^+_{\alpha,{k+\pi}}(0)
f_{\beta,k}(0))>
\end{equation}

So the spin structure factor can be divided into three parts, the 
magnon part ${\cal S}_m$, the hole paet ${\cal S}_h$ and the mixed
part. We can prove that the 
contribution from the mixed term is zero, because the magnon
emission and absorption vertex function satisfy $g(k,0)=-g(-k,0)$.
Using the fluctuation-disipative theorem, we have
\begin{equation}
{\cal S}_m(\pi,\omega)=3\left [ImD_1(0,\omega)-
ImD_1(0,-\omega)\right ]\left [1+n_b(\omega)\right ]
\end{equation}
in which $D_1(0,\omega)$ is the dressed magnon Green's function with
$S_z=1$.
\begin{equation}
{\cal S}_h(\pi,\omega)=Im\chi_h(\pi,\omega)n_b(\omega)
\end{equation} 
in which $\chi_h(\pi,\omega)$ is the spin susceptibility of holes.

Firstly the dressed Green's function of magnons is calculated
by considering the self energy diagram in Fig.7(a,b). Fig.7(a)
represents the contribution of scattering term and the
$\Lambda_s(P,\omega)$ is the effective scattering potential of
the holes and magnons, which can be easily derived from the
full two particle Green's function $\Gamma_s(P,\omega)$ as
$\Lambda_s(P,\omega)=V_s+V_s\Gamma(P,\omega)V_s$ in which
index s represent  two different channels with total
spin 1/2 and 3/2. The Fig.7(b) represent the four kinds
of contribution from the magnon emission and absorption
term which correspond to single particle to single particle,
single particle to two particle, two particle to single particle
and two particle to two particle transition respectively.
And the diagrams considered by us to calculate 
$\chi_h(\pi,\omega)$ are shown is Fig.7(c), which is quite similar
with Fig.7(b) except that there are no vertex function $g(k)$.

Since in this paper we are only interested in the low energy response
within the Haldane gap, we can discard the high energy part during
our calculation. When t is much larger than $t_c$, the band minimum
of the lowest excitation is at zero. Therefore the diagrams in
Fig.7(b,c) describe the intra band transition 
from zero to $\pi$
which costs the energy $4t$. In the large t limit this energy scale
is much larger than the Haldane gap, so the intra band transition
can be ignored in the large t limit. Therefore the only important
diagram in large t limit is Fig.7(a), which is the inter band 
transition from the two particle bound state to the one particle 
state. According to Fig.7(a) the self energy of the magnon can 
be written as,

\begin{equation}
\Sigma(k,\mu,i\omega)=\int dP\sum_{\alpha,m,s,n}G^0(i\nu_n-i\omega,P-k)
D_{\mu,\alpha}^{s,m}\Lambda_s(i\nu_n,P)
\end{equation}
in which $\mu$ and $\alpha$ are the spin index of magnon and hole 
respectively.

$\Lambda_s(i\nu_n,P)$ can be expressed by its spectral function as
\begin{equation}
\Lambda_s(i\nu_n,P)={{\tilde\Lambda_s}\over {i\nu_n-\Delta_H-{P^2\over {2M}}
+{1\over 2}\mu V_s^2}}
+\int d\omega' {{\rho_c(\omega')_c} \over {i\nu_n-\omega'}} 
\end{equation}
in the above expression the first term is contributed by the bound state with
${\tilde\Lambda_s}=\mu V_s^3$ and ${\rho_c(\omega')}$ and the second term which is nonzero
above the Haldane gap is attributed
to the scattering state of one magnon and one hole. Since we are only interested
in the spin response whithin the Haldane gap, we can omit the continuous part and
only keep the first term in the above equation. Then we have
$$
Im\Sigma(\omega,\mu,k)=\int dP\sum_{\alpha,m,s}{\tilde\Lambda_s}D_{\mu,\alpha}^{s,m}\left [
n_F({(P-k)^2\over {2m_f}})-n_F(\Delta_H-{1\over 2}\mu V_s^2-{P^2 \over {2M}})\right ]
$$
\begin{equation}
\delta(\omega-\Delta_H+{1\over 2}\mu V_s^2+{P^2 \over {2m_f}}-{P^2 \over {2M}} ) 
\end{equation}

The spin response function whithin the Haldane gap
 for t=2 (large t limit) is shown in fig.8(a), we can see clearly
that there is a resonate peak in the Haldane gap.
This low energy peak can be attributed to the transition from the bound state
of one magnon and one hole with total spin 1/2 (which is the ground
state under the parameters we chosen here) to the free hole 
state.

The situation of small t limit is quite different with the large
t limit. Now the band minimum of the lower state is moved toward
$\pi/2$. So unlike the large t limit, the holes 
dressed with magnons are distributed near $\pi/2$. Therefore the
intra band transition described by fig.(7b) and fig.(7c) become
very important now, because the momentum transfer $\pi$ costs
very small energy for the states near $\pi/2$. While the energy 
cost for the inter band transition is increased by the virtual
magnon emission and absorption process. So in small t limit
all of the diagrams in fig.7 must be considered.
We have calculated all these diagrams numerically, and the result
is shown in fig.(8b) for $t=0.1\Delta_H$. The contribution from all these figures
are evaluated under the same approximation as the large t limit which
keeps only the low energy peak in the spectral function. For example
the contribution from the second diagram in fig.7(b) can be written as
\begin{equation}
\Sigma_1(i\omega,\pi)=\int dP \sum_{\nu}\gamma^2(P,i\nu,i\omega)
\Lambda_{1\over 2}(P,i\nu)\Lambda_{1\over 2}(P+\pi,i\nu+i\omega)
\end{equation}
in which
$$
\gamma(P,i\nu,i\omega)=\int {-1\over {2\pi\beta}}dk\sum_{\nu_1}
{1\over{i\nu_1-\epsilon(k)}}\cdot{1\over{i\omega+i\nu_1-\epsilon(k+\pi)}}
\cdot{1\over{i\nu-i\nu_1-\Delta_H-{(P-k)^2\over{2m_b}}}} 
$$
$$
=-\int {dk\over {2\pi}}\{ n_F[\epsilon(k)]{1\over{i\omega+\epsilon(k)-\epsilon(k+\pi)}}
\cdot{1\over{i\nu-\epsilon(k)-\Delta_H-{(P-k)^2\over{2m_b}}}} 
$$
$$
+n_F[\epsilon(k+\pi)]{1\over{\epsilon(k+\pi)-i\omega-\epsilon(k)}}
\cdot{1\over{i\nu+i\omega-\epsilon(k+\pi)-\Delta_H-{(P-k)^2\over{2m_b}}}}
$$
\begin{equation}
+\left [1+n_b(\Delta_H+{(P-k)^2\over{2m_b}})\right ]
{1\over{i\nu-\Delta_H-{(P-k)^2\over{2m_b}}-\epsilon(k)}}
\cdot{1\over{i\nu+i\omega-\epsilon(k+\pi)-\Delta_H
-{(P-k)^2\over{2m_b}}}} \}
\end{equation}
whith $\epsilon(k)=-2t[cos(k)-2]$. 
The first two terms in the above equation can be
ignored if we only consider the low temperature case. Also we use the same approximation
that only the low energy peak of $\lambda(P,i\nu)$ is considered.Then after the summation
of $i\nu$ has been done, we have
$$
Im\Sigma_1(\omega,\pi)=\int dP\sum_{mn}\delta(\omega-\omega_m(P+\pi)+
\omega_n(P)){\tilde\Lambda}_n(P){\tilde\Lambda}_m(P+\pi)
$$
\begin{equation}
[n_F(\omega_n(P))
-n_F(\omega_m(P+\pi))]
\gamma^2(P,\omega_n(P),\omega)
\end{equation}
in which $\omega_n(P)$ is the dispersion relation of the two branches obtained in 
equation(42).
The other diagrams in fig.7(b) can be calculated similarly. Finally we calculated
the numerical result for the spin response function in small t limit.
The result is shown in 
Fig.8(b).
We can see clearly from fig.8
that a diffusion like peak is present at low energy near zero
resulted from the intra band transition in small t limit. The 
energy of the inter band transition is near $1.22\Delta_H$ which
is out of the range in which we are interested. 

Our results in large t limit are consistent with the results
of Dagoto and {\it et al} which have two peaks represent
the bound state to single hole state and the intrinsic magnon
excitation respectively. In the paper of Dagoto 
and {\it et al}\cite{Dag}, the dynamically spin structure factor in small
t limit has not been calculated. So our calculation is the first
work which indicate the difference in the dynamically spin 
structure factor for large t limit and small t limit. We
find that when t is smaller than a critical value, the main
contribution to the dynamically spin structure factor will
change from the inter band transition to intra band 
transition.

\bigskip
\newpage
\centerline{\bf Appendix}
In the appendix, we derive the representation of field 
${\vec l}(q)$ and $l^2(q)$.

First we have
$$
{\vec l}(q)=\sum_{k}{\vec\phi}(k+q)\times{\vec\Pi}(-k)
=\sum_{k}{i\over 2}\sqrt{E_k\over E_{k+q}}
({\vec a}_{k+q}+{\vec a}_{-k-q}^+)\times
({\vec a}_{-k}-{\vec a}_{k}^+)
$$
\begin{equation}
=\sum_{k}{i\over 2}\sqrt{E_k\over E_{k+q}}
(-{\vec a}_{k+q}\times{\vec a}_{k}^++
{\vec a}_{-k-q}^+\times{\vec a}_{-k}+
{\vec a}_{k+q}\times{\vec a}_{-k}-
{\vec a}_{-k-q}^+\times{\vec a}_{k}^+)
\end{equation}
It is quite clear that the first two terms describe the 
scattering process and the last two terms describe the
multi-magnon processes which is unimportant when we are
only interested in the low energy physics. So the terms
describing the multi-magnon processes can be omitted
in our present paper. And we have
\begin{equation}
{\vec l}(q)\approx \sum_{k}\gamma_1(k,q)
{\vec a}_{k}^+\times{\vec a}_{k+q}
\end{equation}
in which $\gamma_1(k,q)=i\sqrt{E_k\over E_{k+q}}$.

$$
l^2(q)=\sum_k{\vec l}(k+q)\cdot {\vec l}(-k)
=\sum_{k,k',k''}\gamma_1(k',k+q)\gamma_1(k'',-k)
({\vec a}_{k'}^+\times{\vec a}_{k+k+q'})\cdot
({\vec a}_{k''}^+\times{\vec a}_{k''-k})
$$

$$
=\sum_{k,k',k''}\gamma_1(k',k+q)\gamma_1(k'',-k)\cdot [
-2a_{y,k'}^+a_{y,k''-k}\delta_{k'+k+q,k''}
-2a_{x,k'}^+a_{x,k''-k}\delta_{k'+k+q,k''}
-2a_{z,k'}^+a_{z,k''-k}\delta_{k'+k+q,k''}
$$

$$
+a_{x,k'}^+a_{x,k''}^+a_{y,k'+k+q}a_{y,k''-k}
+a_{y,k'}^+a_{y,k''}^+a_{z,k'+k+q}a_{z,k''-k}
+a_{z,k'}^+a_{z,k''}^+a_{x,k'+k+q}a_{x,k''-k}
$$

$$
+a_{y,k'}^+a_{y,k''}^+a_{x,k'+k+q}a_{x,k''-k}
+a_{z,k'}^+a_{z,k''}^+a_{y,k'+k+q}a_{y,k''-k}
+a_{x,k'}^+a_{x,k''}^+a_{z,k'+k+q}a_{z,k''-k}
$$

$$
-a_{y,k'}^+a_{x,k''}^+a_{x,k'+k+q}a_{y,k''-k}
-a_{z,k'}^+a_{y,k''}^+a_{y,k'+k+q}a_{z,k''-k}
-a_{x,k'}^+a_{z,k''}^+a_{z,k'+k+q}a_{x,k''-k}
$$

$$
-a_{x,k'}^+a_{y,k''}^+a_{y,k'+k+q}a_{x,k''-k}
-a_{y,k'}^+a_{z,k''}^+a_{z,k'+k+q}a_{y,k''-k}
-a_{z,k'}^+a_{x,k''}^+a_{x,k'+k+q}a_{z,k''-k} ]
$$
After discarding the multi-magnon terms, we have
\begin{equation}
l^2(q)=\sum_{k',\alpha}2\gamma'(k',q)a_{\alpha,k'}
^+a_{\alpha,k'+q}
\end{equation}
with $\gamma'(k',q)=\sum_k\gamma_1(k',k+q)\gamma_1(k'+k+q,-k)$.

\newpage

\bigskip
\centerline{\bf Figure Caption}
\vspace{.5cm} 

Fig.1 The illustration of holes doped antiferromagnetic chain.

Fig.2 The camparison of our result and the numerical result for
the lowest excitation in static hole limit. The line represents our
result and the squares represent the corresponding numerical result.

Fig.3 The Feyman diagrams used in this paper. (a) The full line represents
the hole's Green Function. (b)The dashed line represents the magnon's
Green function. (c) The scattering vortex contributed by $H_1$.
(d) The magnon emission and absorption vortex contributed by $H_1$.

Fig.4 (a) The dressed two-particle Green function with total spin 3/2.
(b) The dressed two-particle Green function with total spin 1/2.
(c) the dressed hole's Green function.

Fig.5 From down to up,
 the up three curves represent the $\omega_+(P)$ for 
 t=0.22,0.20,0.18 respectively. And from up to down, the lower three
 curves represent the $\omega_-(P)$ for 
 t=0.22,0.20,0.18 respectively.
 
Fig.6 The full and dashed lines represent the spectral weight $z_+(P)$
and $z_-(P)$ respectively.

Fig.7 (a) The magnon self energy caused by the scattering term.
(b) The magnon self energy caused by the magnon emission and absorption
 term. (c) The hole's contribution to the spin susceptibility.
 
Fig.8 (a) The spin structure factor in large t limit (t=2).
(b)The spin structure factor in samll t limit (t=0.1).


\begin{references}
\vspace{1.0ex}
\baselineskip=14pt

\bibitem{Hal} F. D. M. Haldane, Phys. Lett. {\bf 93A}, 464(1983);
Phys. Rev. Lett. {\bf 50}, 1153 (1983).

\bibitem{exp} W. J. L. Buyers, R. M. Morra, R. L. Armstrong, 
M. J. Hogan, P. Gerlach, and K. Hirikawa, Phys. Rev. Lett.
{\bf 56}, 371 (1986);
J. P. Renard, M. Verdaguer, L. P. Regnault, W. A. C. Erkelens,
J. Rossat-Mignod, and W. G. Stirling, Europhys. Lett. {\bf 3},
945 (1987).

\bibitem{th1}R. Botet and R. Julien, Phys. Rev. B {\bf 27}, 613
(1983);
M. P. Nightingale and H. W. J. Blote, Phys. Rev. B {\bf 33}, 659
(1986);
H. J. Schultz and T. A. L. Ziman, Phys. Rev B {\bf 33}, 6545 (1986).

\bibitem{th2} I. Affleck, T. Kennedy, E. H. Lieb, and H. Tasaki,
Phys. Rev Lett. {\bf 59}, 799 (1987).

\bibitem{DMRG} S. R. White and D. A. Huse, Phys. Rev. B {\bf 48}, 
3844 (1993).

\bibitem{ext} O. Golinelli, Th. Jolicoeur, and R. Lacaze, Phys. Rev. B
{\bf 50}, 3037 (1994).

\bibitem{NLS} I. Affleck, Field Theory Methods and Quantum Critical
Phenomena, edited by E. Brezin and J. Zinn-Justin (North-Holland,
Amsterdam, 1990).

\bibitem{FB1} I. Affleck, Phys. Rev. Lett. {\bf 62}, 474 (1989).

\bibitem{FB2} I. Affleck and G. F. Wellman, Phys. Rev. B {\bf 46}
8934 (1992).

\bibitem{FB3} J. Sagi and I. Affleck, cond-mat/9506137.

\bibitem{exp2} D. J. Buttrey, J. D. Sullivan, and A. B. Rheingold,
J. Solid State Chem. {\bf 88}, 291 (1990);
J. Amador {\it et al}, Phys. Rev. B {\bf 42}, 7918 (1990);
R. Saez-Puche {\it et al}, J. Solid State Chem. {\bf 93}, 461(1991);
J. Darriet and L. P. Regnault, Solid State Commun. {\bf 86}, 409
(1993).

\bibitem{exp3} B. Batlogg, S-W. Cheong, and L. W. Rupp, Jr., Physica
(Amsterdam) {\bf 194B-196B}, 173 (1994);
A. P. Ramirez {\it et al}, Phys. Rev. Lett. {\bf 72}, 3108 (1994).

\bibitem{dop} J. F. DiTusa {\it et al}, Phys. Rev. Lett. {\bf 73},
1857 (1994).

\bibitem{st1} E. S. Sorensen and I. Affleck, Phys. Rev. B {\bf 51},
16115 (1995).

\bibitem{st2} Z. Y. Lu, Z. B. Su and L. Yu, Phys. Rev. Lett. {\bf 74},
4297 (1995).

\bibitem{st3} M. Kaburagi, I. Harada, and T. Tonegawa, J. Phys. Soc.
Jpn. {\bf 62}, 1848 (1993);
M. Kaburagi and T. Tonegawa, J. Phys. Soc. Jpn. {\bf 63}, 420 (1993).

\bibitem{zhang} S. C. Zhang and D. P. Arovas, Phys. Rev. B {\bf 40},
2708 (1989).

\bibitem{penc} K. Penc and H. Shiba, Phys. Rev. B. {\bf 52}, R715
(1995).

\bibitem{Dag} E. Dagotto, J. Riera, A. Sandvik, and A. Moreo,
Phys. Rev. Lett. {\bf 76}, 1731 (1996).

\bibitem{open1} T. Kennedy, J. Phys. Condens. Matter {\bf 2},
5737 (1990).

\bibitem{open2} S. Miyashita and S. Yamamoto, Phys. Rev. B {\bf 48},
913 (1993);
S. R. White, Phys. Rev. Lett. {\bf 68}, 3487 (1992).


\end{references}
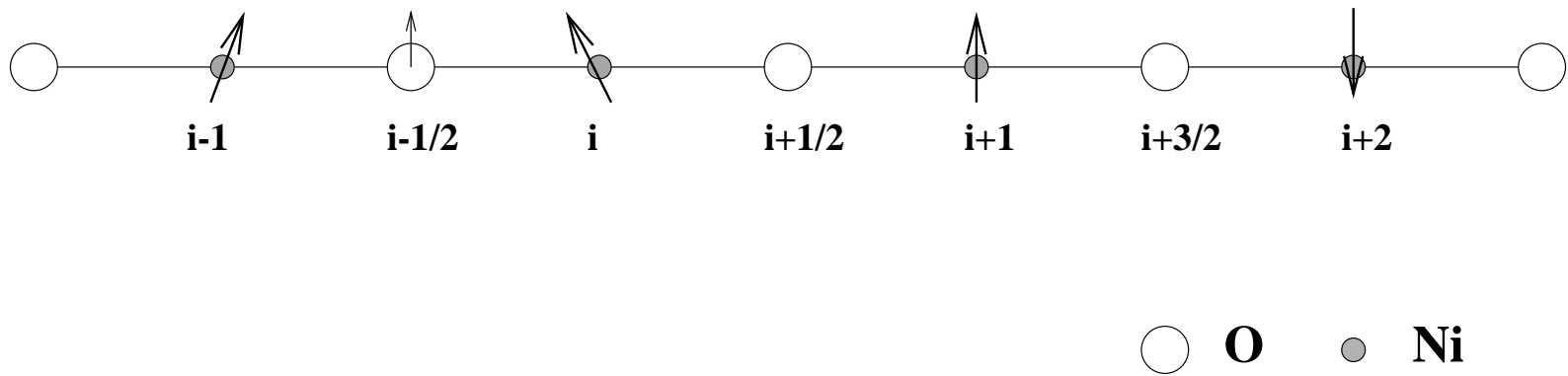
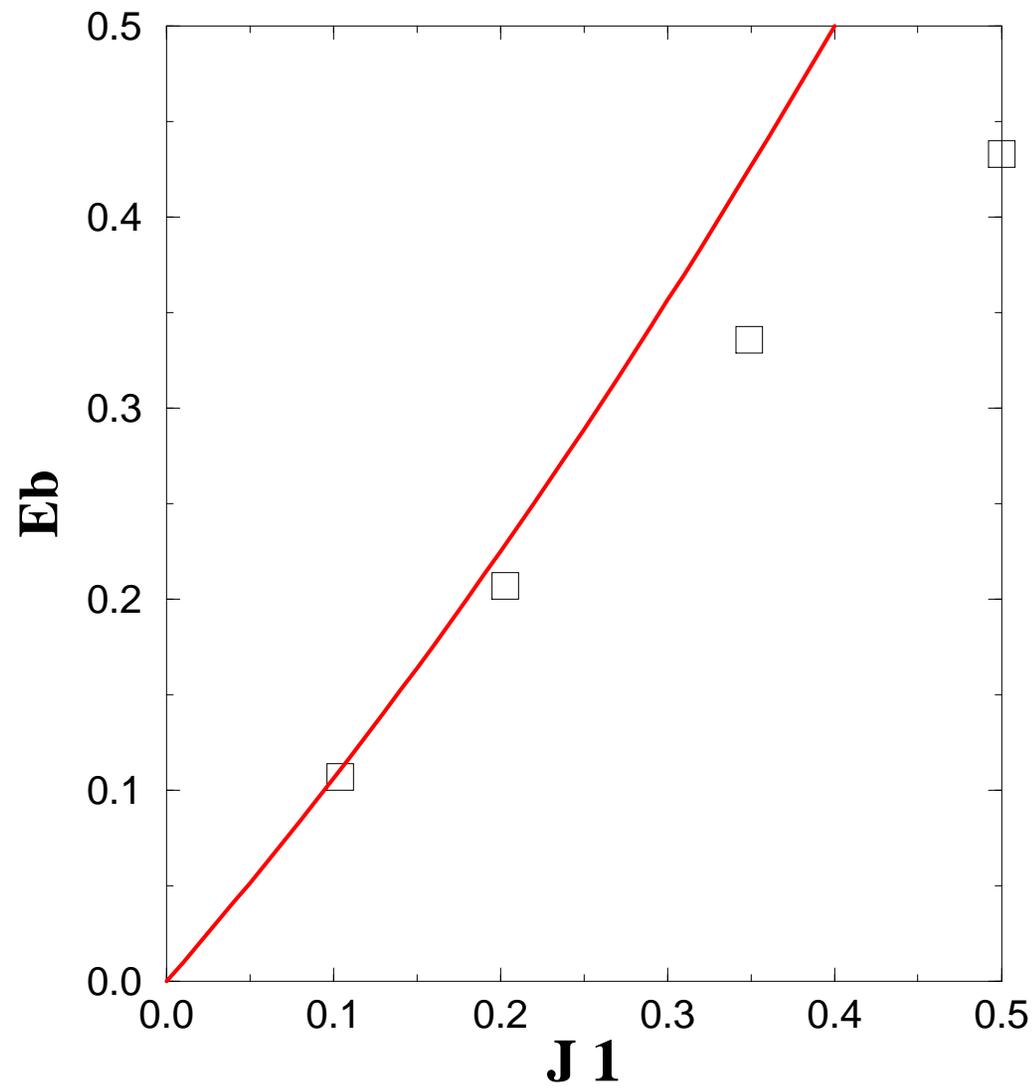
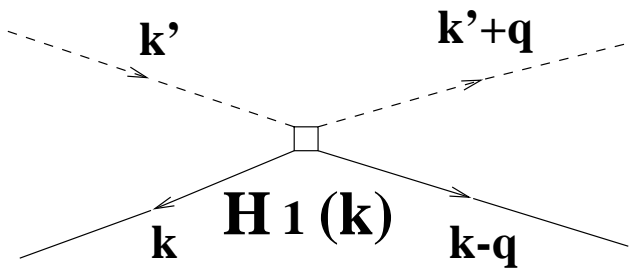
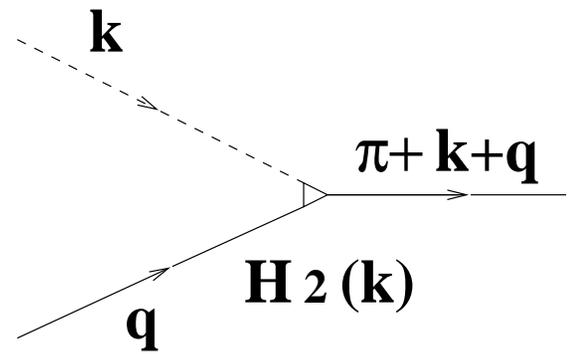
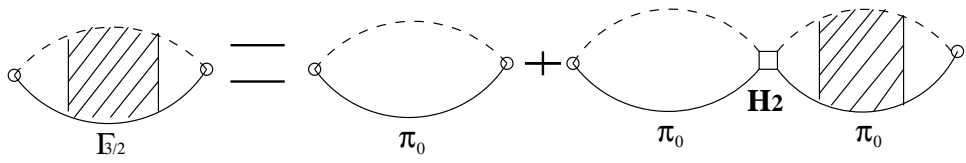
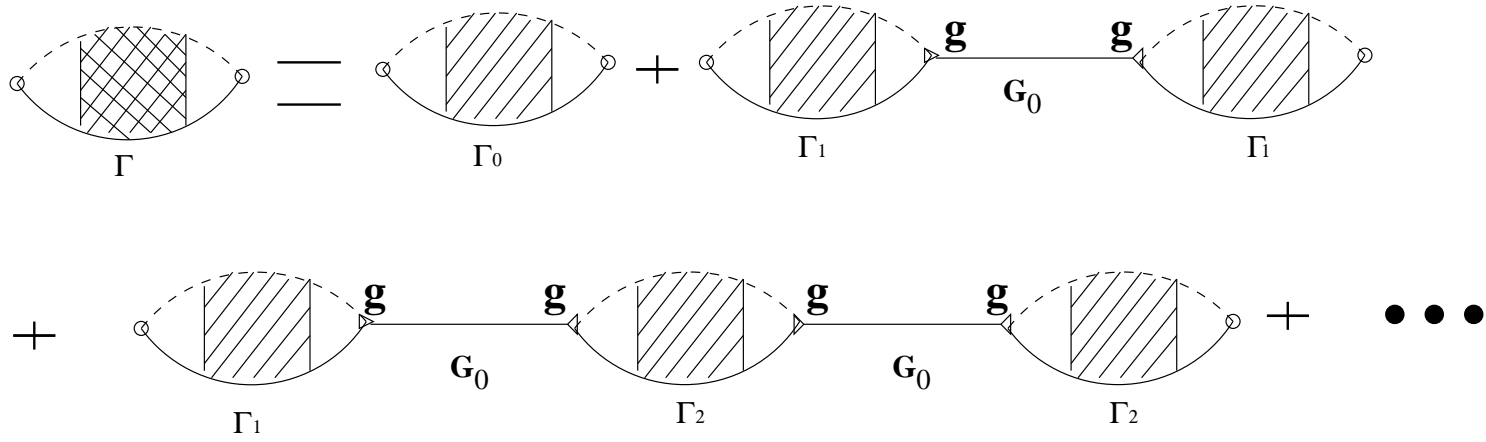
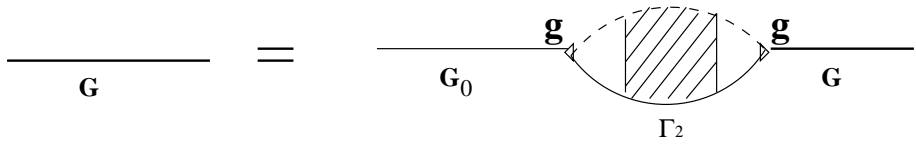
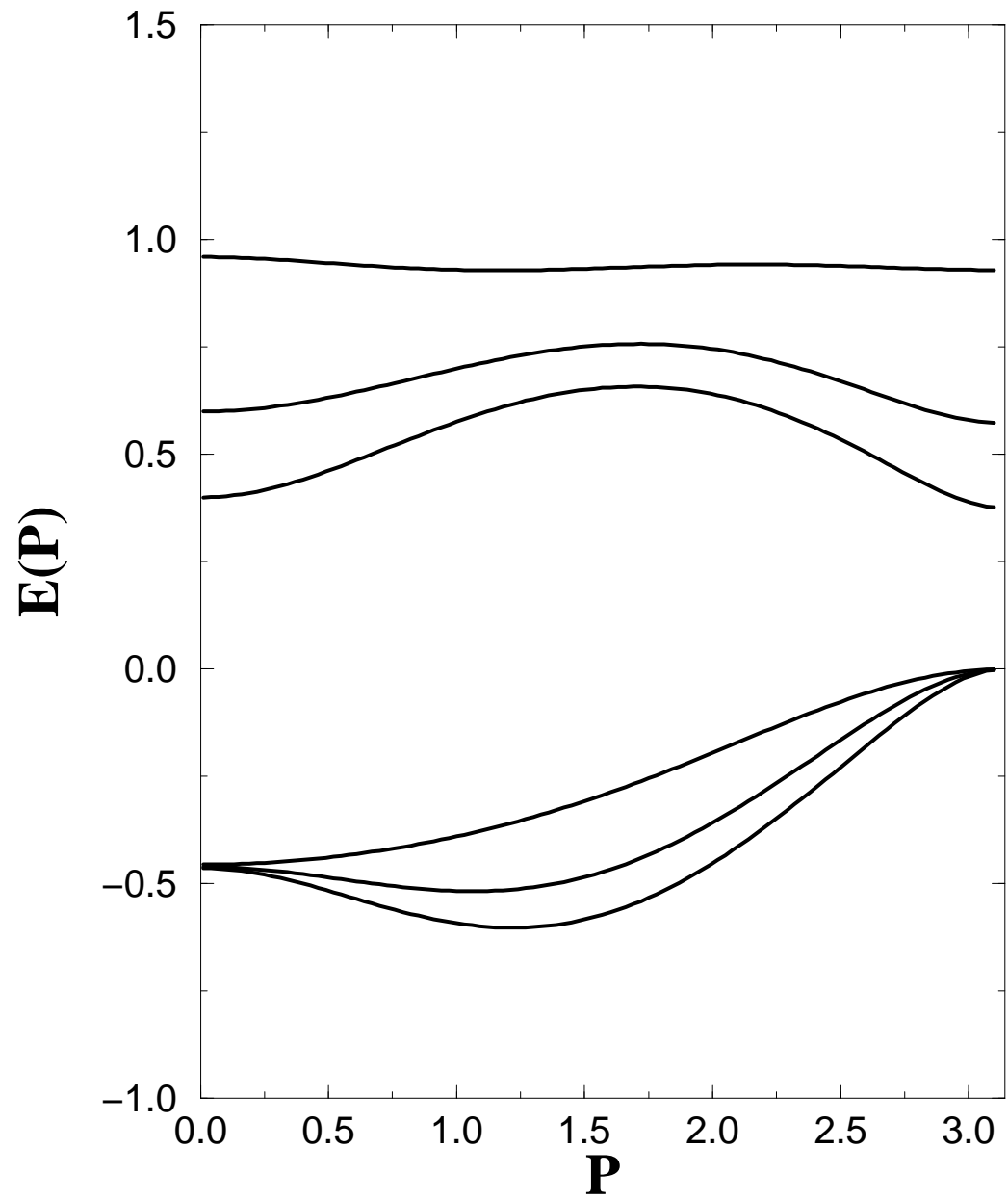
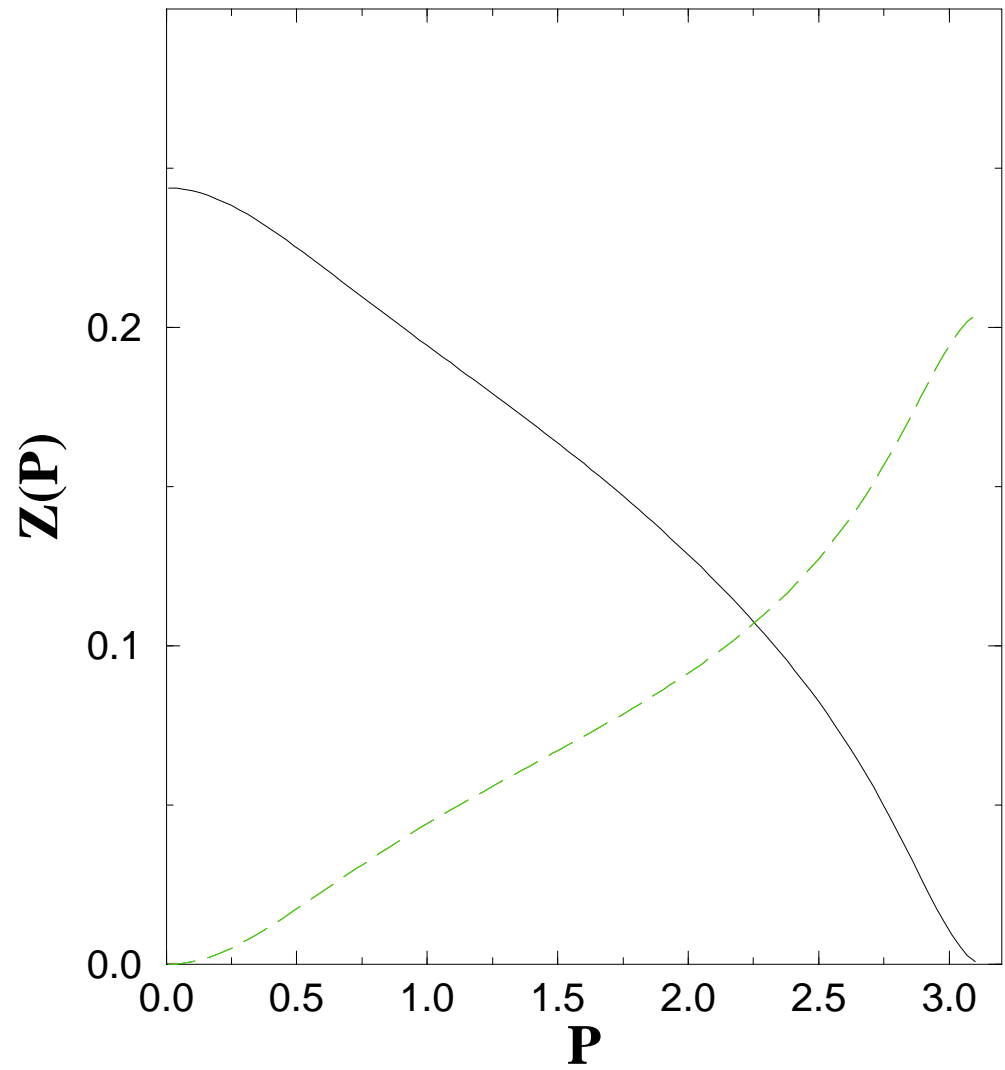
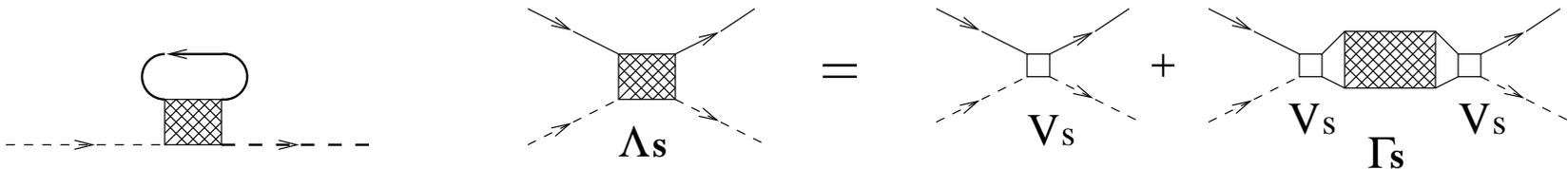
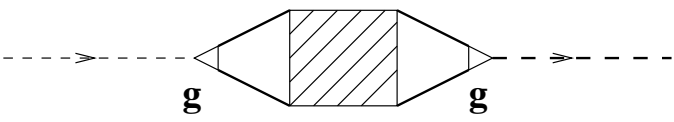
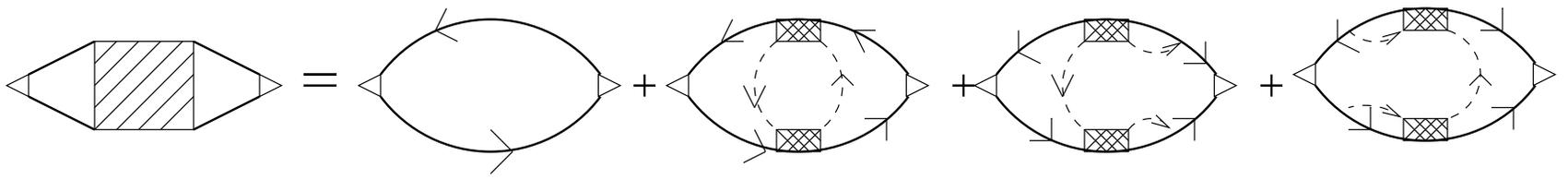
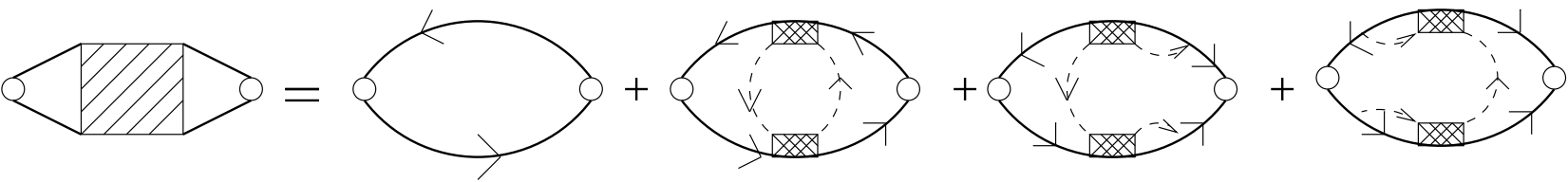
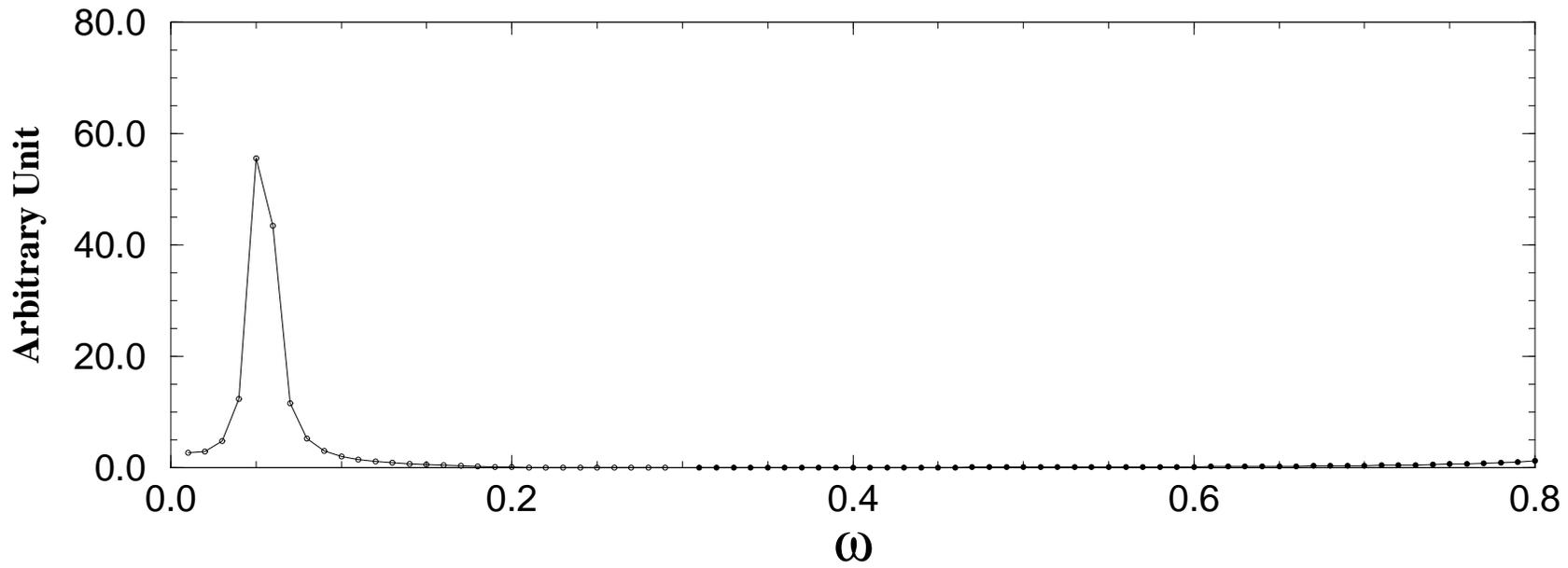
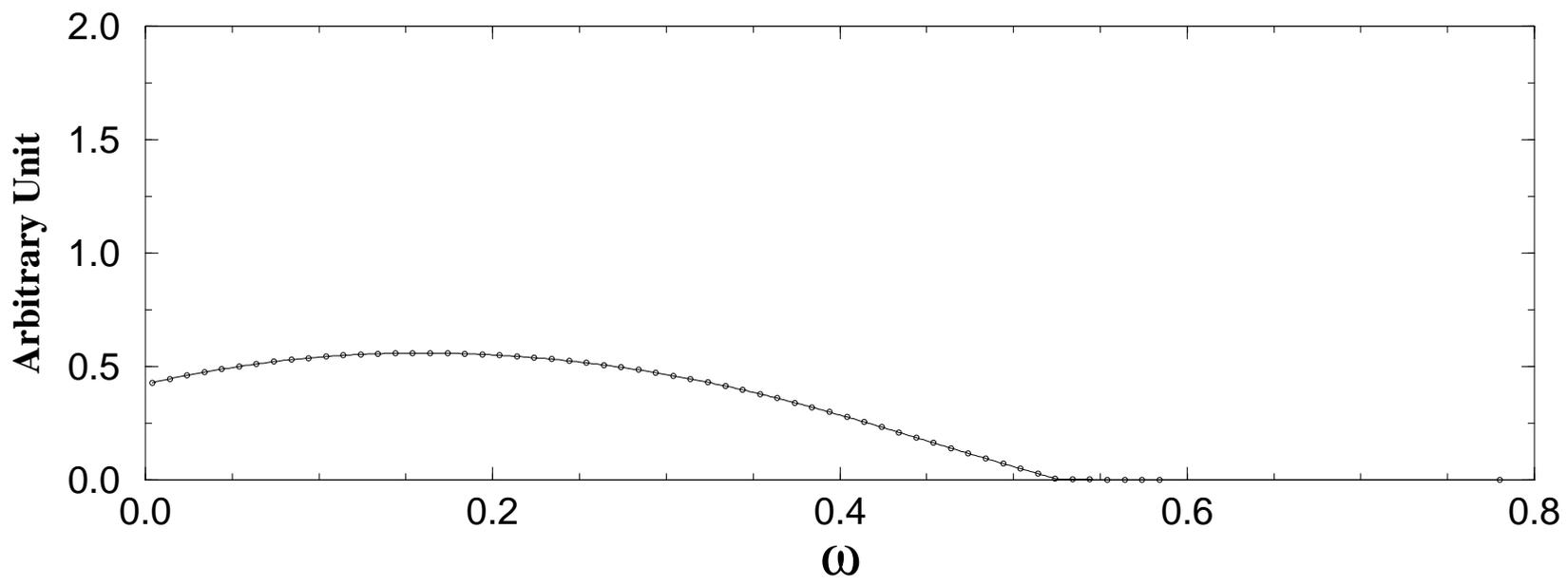
\end{document}